# Temperature-driven evolution of hierarchical nanodomain structure in tetragonal-like BiFeO$_3$ films

*Yajun Qi,[1,2] Zuhuang Chen,[1] Lihua Wang,[3] Xiaodong Han,[3] Junling Wang,[1] Thirumany Sritharan,[1] and Lang Chen [1, a)]*

[1] School of Materials Science and Engineering, Nanyang Technological University, Singapore 639798, Singapore

[2] Department of Materials Science and Engineering, Hubei University, Wuhan 430062, P. R. China

[3] Institute of Microstructure and Properties of Advanced Materials, Beijing University of Technology, Beijing, 100022, P. R. China

**Abstract**

Transmission electron microscopy study of tetragonal-like BiFeO$_3$ films reveals a hitherto unreported hierarchical nanodomain structure. The 30-50 nm wide stripe domains with {110} domain walls consist of a substructure of lamellar nanodomains of 8-10 nm width in a herringbone-like arrangement. *In situ* heating and cooling reveals a reversible transition from the hierarchical nanodomain structure to a tweed-like domain structure which is accompanied by a first-order phase transition near 120 °C with a thermal hysteresis.

[a)] Author to whom correspondence should be addressed; electronic mail: langchen@ntu.edu.sg

Multiferroic BiFeO$_3$ (BFO) has attracted great attention due to the intriguing fundamental physics and the potential for widespread applications [1]. Recently, there has been particular interest in the crystal [2-8] and domain structures [9,10] of the tetragonal-like (T-like) BFO thin films as well as their relationships to electric and magnetic properties [11-15]. The T-like BFO, which can be stabilized by strain, was theoretically predicted to have a giant polarization value of about 150 $\mu$C/cm$^2$ [16]. In our previous reports of piezoelectric force microscopy (PFM) investigations [3,4], we showed the formation of nanoscale stripe domains in T-like BFO films and indicated that this T-like phase is actually a monoclinic $M_C$ phase with polarization vector in the (010) plane. Furthermore, a recent report [14] suggests that the T-like, monoclinic BFO is antiferromagnetic with a predominance of *G*-type character ($T_N$ = 324 K) and a minority of *C*-type character ($T_N$ = 260 K). It showed a transition from hard to soft ferroelectric behavior accompanied by a transition from an antiferromagnetic to a paramagnetic order at 360 ± 20 K [15]. These property transitions were attributed to an observed phase change from the $M_C$ to T-like $M_A$ phase [17]. However, this phase change and the associated domain evolution were not well studied. Transmission electron microscopy (TEM), on the other hand, has a very high spatial resolution compared to PFM, and thus can provide accurate structure information at nanometer scale, particularly with the convergent beam electron diffraction (CBED) technique. In this Letter, we present results of our TEM study of T-like, $M_C$ phase BFO films. We report a hierarchical domain structure in these films and observe its thermally induced evolution by *in situ* heating/cooling.

BFO films of thickness ~50 nm were deposited on (001) LaSrAlO$_4$ (LSAO) single crystal substrates by pulsed laser deposition [3,4]. Plan view TEM specimens were prepared by the standard procedures of cutting, gluing, mechanical polishing, and ion milling except that polishing and milling were done on the side of the substrate only. TEM investigations were carried out on JEOL electron microscopes JEM2100F (FEG) and JEM2010 operated at 200 kV. The CBED technique with a probe size of 2.4 nm was used to confirm the $M_C$ phase of the BFO film. The selected area electron diffraction (SAED) and CBED patterns are indexed on the basis of the pseudo-cubic unit cell. The *in situ* heating/cooling observation was performed on a double-tilt heating specimen holder (Gatan Model 652) with the temperature controlled precisely by a SmartSet Hot Stage controller (Gatan Model 901). The accuracy of the temperature measurement is about 0.1 °C. A heating and cooling rate less than 3 °C/min was used. Images and CBED patterns were recorded 10 minutes after the temperature was stabilized.

Figure 1 shows a bright-field image of the plan view of BFO film taken a few degrees off the [001] zone axis in order to probe the nanodomains clearly. The stripe structure noted previously via PFM in many studies [3,4,9,10] is clear in this TEM image also. The stripe widths are 30-50 nm, and the stripes lie along two perpendicular directions. The SAED pattern obtained from an area of several stripes is shown in the inset of Fig. 1. This figure shows splitting of diffraction spots typical of twin-like boundaries with small tilt angles, as in ferroelectric domain boundaries. From a defocus diffraction pattern, the stripe domain walls were observed to direct along [110] and [1$\bar{1}$0] directions, which is consistent with our previous PFM study [3,4]. More interestingly, a substructure can be detected within each stripe indicating a nanostructure finer than the main stripes. The enlarged image of Fig. 2(a) shows this substructure clearly. It appears to consist of lamellae of nanodomains of width of 8-10 nm branching from a "backbone" of a large stripe domain boundary giving a "herring-bone" appearance. The projections of the traces of the lamellar nanodomain walls in the substructure do not appear to orient in any principle crystallographic directions, but are inclined about 20-30 degree to the "backbone" stripe domain wall. The substructure nanodomains contrasts are not Moiré fringes, which should have an exactly or nearly equal spacing with a fixed orientation with respect to the forming planes [18]. Furthermore, the calculated spacings of the Moiré fringes are much larger than the observed nanodomain width [19]. Thus, the film structure consists of a substructure of lamellar nanodomains of width 8-10 nm in a "herring-bone"-like arrangement within crystallographically oriented larger stripes of width 30-50 nm. Fig. 2(b) shows a simple schematic model of this hierarchical domain structure. A similar hierarchical domain structure was also observed in BFO films grown on LaAlO$_3$ substrate. The high resolution TEM image of Fig. 2(c) taken across a substructure wall indicated in Fig. 2(a) confirms the absence of any deviation of lattice planes across the nanodomain walls. The in-plane lattice parameters deduced from Fig. 2(c) are $a$ = 3.80 Å and $b$ = 3.76 Å which are close to our previous x-ray diffraction data [4].

Figures 3 (a) and (b) shows a typical [001] zone axis CBED patterns of two adjacent lamellar nanodomains within one stripe domain. Similar patterns were obtained at many different locations when adjacent lamellae nanodomains were examined. The patterns exhibit (010) and (100) mirror planes (marked "m"). It was previously shown that a ($1\bar{1}0$) mirror plane is permitted in $M_A$ (or $M_B$) phase while mirror planes of the {100} family could occur in the $M_C$ and tetragonal phases [20]. Thus, we conclude that our BFO film consists of $M_C$ phase since the tetragonal phase is unlikely to form at the misfit strain of ~−5% that exist between BFO and the LSAO substrate [5, 6]. One also note that the LSAO substrate may not be completely removed in the relatively thick area of plan-view TEM sample and thus is still able to stabilize the $M_C$ phase.

Similar hierarchical nanodomain structures have been observed in (1−x)Pb(Mg$_{1/3}$Nb$_{2/3}$)O$_3$-xPbTiO$_3$ (PMN-PT) single crystals and Pb(Zr$_{1−x}$Ti$_x$)O$_3$ (PZT) solid solution at chemical compositions near their morphotropic phase boundaries (MPB) [20-22]. For example, in PMN-PT, besides a intrinsic monoclinic $M_C$ structure reported by Noheda *et al.* [23], others suggested that the monoclinic phase could be an adaptive phase [24,25] where the tetragonal *c/a* nanodomains assembled together into submicrodomains, and thus the overall microdomain exhibited a monoclinic nature due to an averaging effect [20,24,25]. In our case, the $M_C$ phase of BFO, has a hierarchical nanodomain structure, but consists of the same $M_C$ phase in both the stripe domains and the lamellar substructure, which is unlike the adaptive structure of the PMN-PT single crystals. Furthermore, the large tetragonal ratio (*c/a* = 1.25) in T-like $M_C$ phase of BFO wouldn't encourage *a* domains at large compressive stresses and we did not detect the *a/c/a/c…* type polydomain structure, which is reportedly in the adaptive domain picture proposed in cubic-like PMN-PT and PZT near their MPB compositions [20,24,25]. The nanodomain walls in PMN-PT and PZT are along <110> or <100> direction [20,22] while, in BFO, the substructure nanodomain walls do not appear to have any principle crystallographic orientation. This may due to the intrinsic richness of probable low symmetry phases in BFO which has been demonstrated theoretically and experimentally [5,6,8]. The small energy difference between the various low symmetry phases in BFO could favor the assembly of hierarchical nanodomain structures as the lattice can easily accommodate local strains by the variety of low-symmetry phases [5,6].

The important ferroelectric, piezoelectric and magnetoelectric coupling properties of BFO critically depend on the prevailing domain structure [1]. Next we demonstrate a near-room temperature domain structure evolution in BFO films by *in situ* TEM heating and cooling experiments. *In situ* observations were carried out with a heating specimen holder on the JEM 2010. Figs. 4(a)-(g) show serial dark-field TEM images of the same area during the heating/cooling processes. No significant change in domain morphology occurred until 91 °C. Thereafter, at 104 °C the herring-bone nanodomains started to disappear and the domain contrast changed quickly. When 109 °C was reached, the stripe contrast was progressively replaced by a nanosized "tweed-like" contrast. The stripe contrasts completely disappeared at 129 °C, as shown in Fig. 4(d). No further changes in contrast were noted with increasing temperature up to 450 °C suggesting that the ferroelectric to paraelectric phase transition temperature should be higher than 450 °C. The [001] zone axis CBED pattern obtained at 130 °C showed ($1\bar{1}0$) mirror plane as in Fig. 4(h). This is a characteristic feature of the $M_A$ and rhombohedral phases [20] signifying that the "tweed-like" domains are from $M_A$ or rhombohedral phase. A complimentary parallel investigation carried out on temperature-dependent, grazing incidence x-ray diffractometer showed that the periodic $M_C$ domains begin to disappear at ~90 °C and the $M_C$ phase transforms to T-like $M_A$ phase for temperatures higher than 100 °C [26]. A very recent study by Siemons *et al*. also suggested that the higher temperature phase is a $M_A$ phase from the temperature-dependent x-ray diffraction data [17]. These evidences further support that the domain configuration changes correspond to the structural transition from $M_C$ to $M_A$ phase as the temperature is increased from room temperature.

No changes were detected on cooling from 450 °C until 80 °C when a wave-like contrast with an average size of 40 nm appeared, as evident in Fig. 4(e). This gradually gave way to the stripe contrast containing herring-bone structure at 68 °C (Fig. 4(f)). The stripes were found to be along <110> direction. When cooled to 57 °C, the tweed-like and wave-like contrasts disappeared completely leaving only the stripe contrast with herring-bone hierarchical structure of the original sample. This proves the reversibility of the phase changes and their corresponding domain patterns (Fig. 4(g)). The difference in the transformation temperatures observed during heating and cooling cycles indicates a thermal hysteresis. Thus, this domain pattern change and the related structural change might be a first-order phase transition with a large hysteresis temperature range.

In summary, the 8-10 nm herring-bone-like nanodomains were observed in T-like $M_C$ phase BFO films by TEM. The nanodomains assembled into stripe domains forming a hierarchical structure and the stripe domain walls appear to be the backbone of the herring-bone structure. The reversible domain transition corresponding to a structural phase transition near 120 °C was shown by *in situ* heating and cooling in a TEM. It is possible that these ultrafine, lamellar nanodomains could easily respond to an external field and the cooperative response and may give rise to outstanding piezoelectric properties. It is also expected that the structural phase transitions close to room temperature lead to modified magnetic and ferroelectric properties and would open the door towards new application in the model multiferroic $BiFeO_3$.

**ACKNOWLEDGEMENT**

We acknowledge the support from the Singapore National Research Foundation under its CREATE program: Nanomaterials for Energy and Water Management, the support from MINDEF-NTU-JPP 10/12 and ERI@N.

**Figure captions**

Fig. 1 (Color online) Plan-view bright-field TEM image of the BFO film taken near the [001] zone axis. Inset: [001] zone axis SAED pattern taken from the area with several stripes.

Fig. 2 (Color online) (a) High magnification of the hierarchical nanodomain structure. (b) A schematic of hierarchical domain configuration on the (001) plane. (c) High resolution TEM image near the nanodomain wall area indicated by a circle in (a).

Fig. 3 (Color online) [001] zone axis CBED patterns of the adjacent lamellar nanodomains within the same stripe domain. "m" denote the mirror plane. (010) and (100) mirror planes appear in pattern (a) and (b), respectively.

Fig. 4 (Color online) *In situ* observation of the same area during heating/cooling: (a) natural hierarchical domains at 38°C, (b) 104 °C on heating, (c) 109 °C on heating, (d) 129 °C on heating, (e) 72 °C cooling from 450 °C, (f) 68 °C on cooling, and (g) 25 °C on cooling; (h) [001] zone axis CBED pattern of the tweed-like domain obtained at 130 °C.

Fig. 1

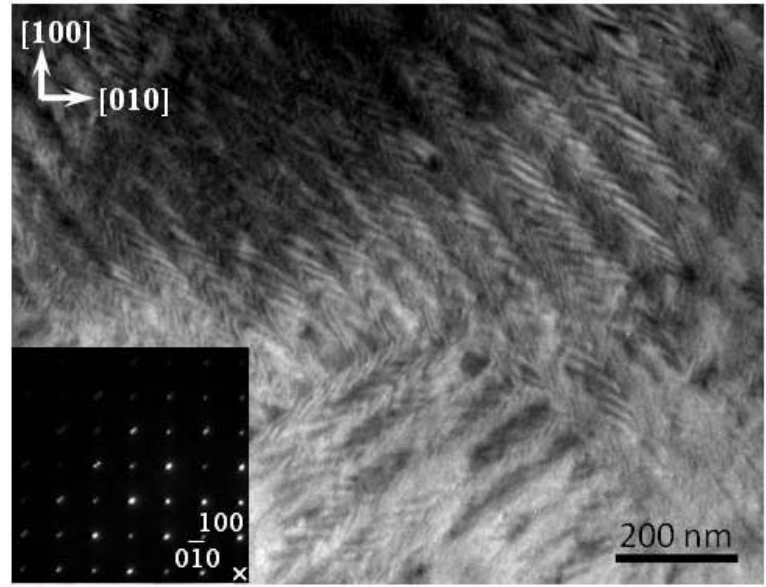

Fig. 2

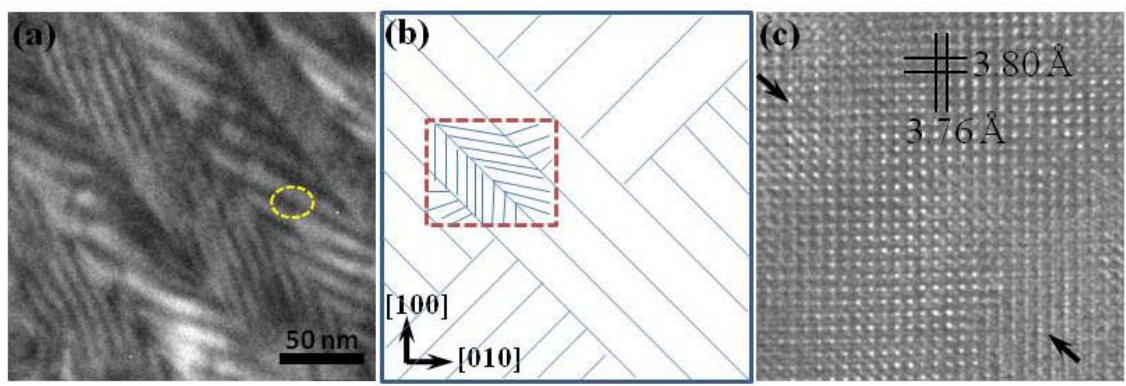

Fig. 3

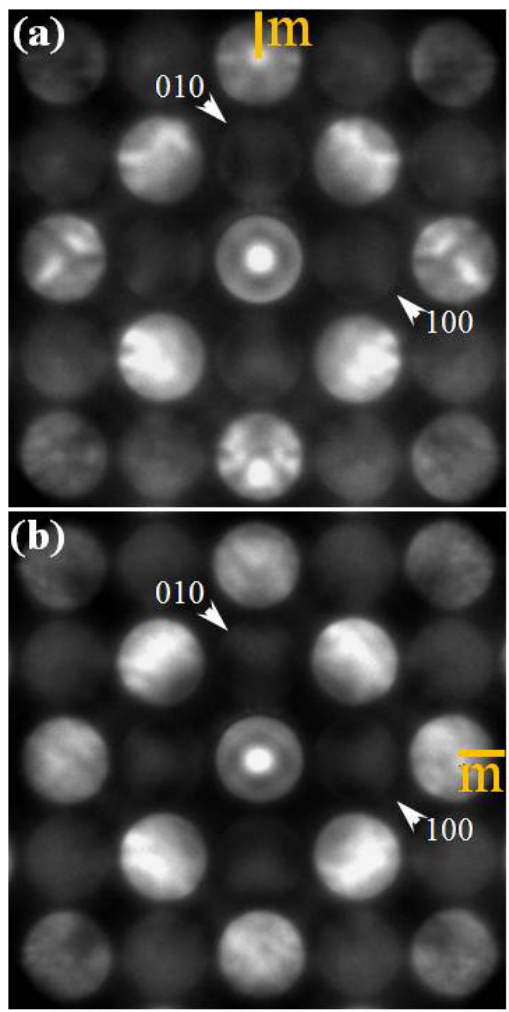

Fig. 4

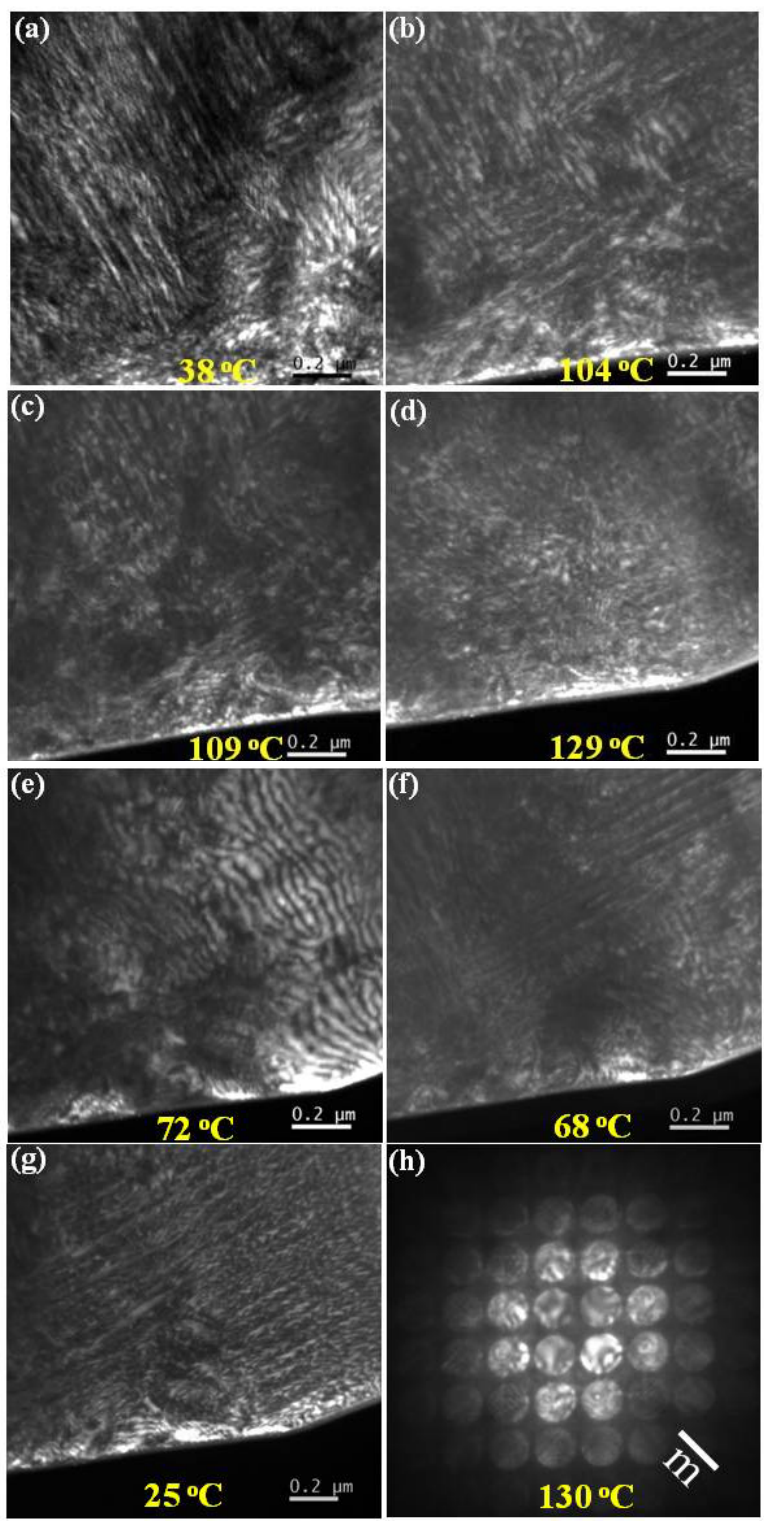